\documentclass[journal=nalefd,manuscript=letter,layout=twocolumn]{achemso}

\usepackage[version=3]{mhchem} 
\usepackage{amsmath}

\usepackage{hyperref}
\hypersetup{colorlinks=true, linkcolor=blue, anchorcolor=blue, citecolor=blue, urlcolor=blue}

\usepackage{multirow}

\newcommand{\mobunit}{cm$^2$V$^{-1}$s$^{-1}$}
\author{Zirui He}
\affiliation[1]{Department of Materials Science, Fudan University, Shanghai 200433, China}
\alsoaffiliation[2]{Shanghai Advanced Silicon Technology Co., Ltd., Shanghai 201616, China}
\author{Shang-Peng Gao}
\email{gaosp@fudan.edu.cn}
\affiliation[1]{Department of Materials Science, Fudan University, Shanghai 200433, China}
\alsoaffiliation[2]{Shanghai Advanced Silicon Technology Co., Ltd., Shanghai 201616, China}
\author{Meng Chen}
\affiliation[2]{Shanghai Advanced Silicon Technology Co., Ltd., Shanghai 201616, China}
\email{mchen@ast.com.cn}

\title{First-principles Prediction of Carrier Mobility in Semiconductor Nanowires Based on the Spatially Dependent Boltzmann Transport Equation}

\keywords{carrier mobility, electron-phonon scattering, surface scattering, nanowire, Boltzmann transport equation, first-principles calculation}

\begin{document}

\begin{tocentry}

\includegraphics[width=\linewidth]{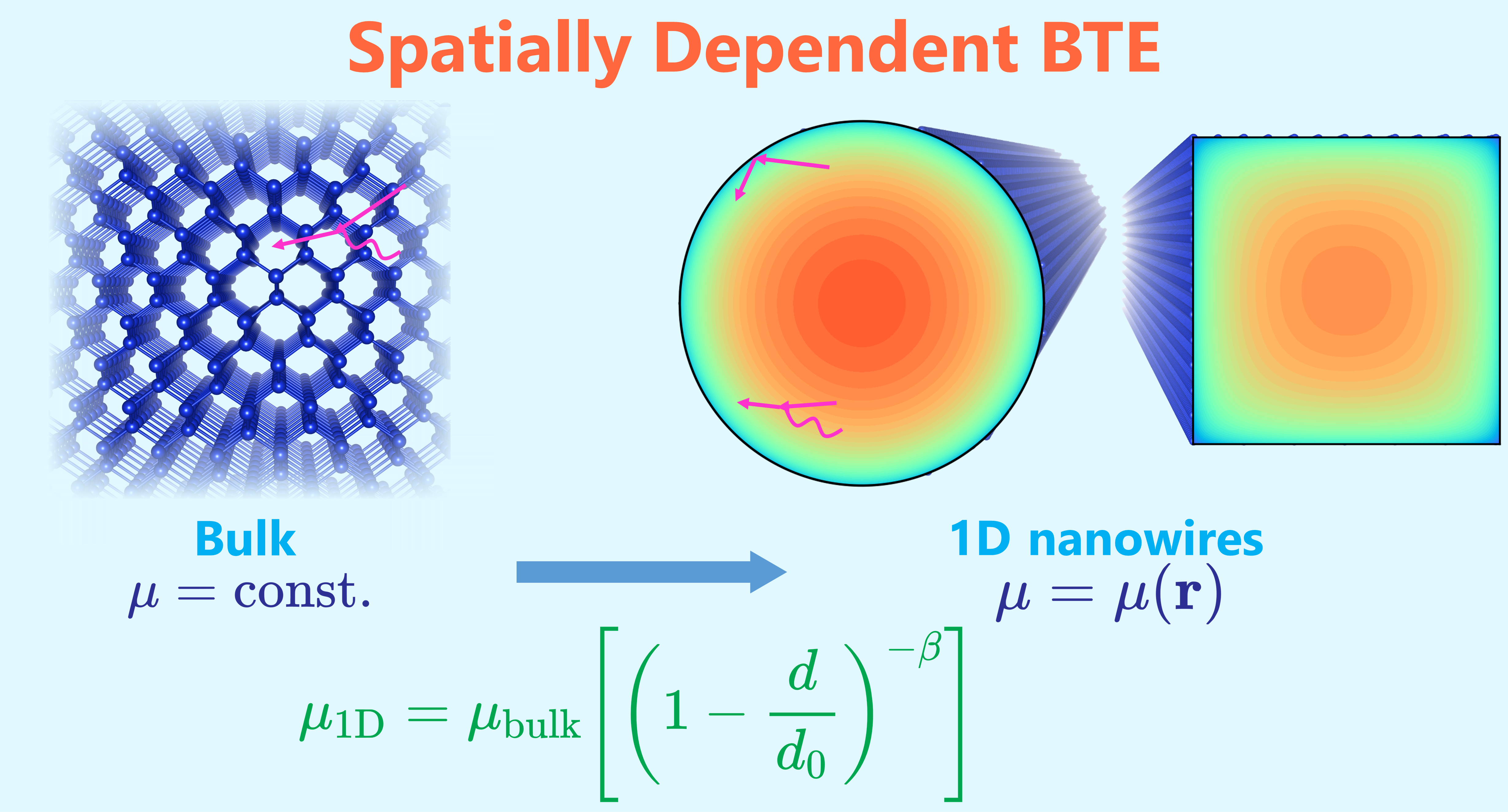}





\end{tocentry}


\begin{abstract}
  Carrier mobility in bulk semiconductors is typically governed by electron-phonon (e-ph) scattering.
  In nanostructures, spatial confinement can lead to significant surface scattering, lowering mobility and breaking the spatial homogeneity assumption of conventional models.
  In this work, a fully \latin{ab initio} framework based on the spatially dependent Boltzmann transport equation for one-dimensional nanowires is developed.
  We apply it to Si and GaN assuming diffusive surface scattering, and reveal the mobility-diameter relation: $\mu_\mathrm{1D} = \mu_\mathrm{bulk} \left[1-\left(d/d_0\right)^{-\beta}\right]$.
  The parameter $d_0$, comparable to the carrier mean free path, defines a boundary layer exhibiting a considerable mobility gradient, and also quantifies the competition between e-ph and surface scattering together with $\beta$.
  We further discuss the effects of orientation, cross-sectional shape, and temperature.
  Moreover, experimental data are generally lower than our predictions, possibly due to structural imperfections, systematic errors from measurements, \latin{etc}.
  Therefore, our theoretical method can provide an intrinsic benchmark toward optimized experimental realizations.
\end{abstract}

\vspace{1cm}

Carrier mobility is a basic parameter that determines the efficiency of charge transport in semiconductors, with a fundamental effect on various properties from electronic conductivity to device frequency response.
Electron-phonon (e-ph) interaction is an intrinsic factor that affects carrier mobility, and considerable effort has been devoted to enable a quantitative description of it.
The fully \latin{ab initio} methodology developed in recent years can provide enlightening insights into the microscopic e-ph scattering mechanisms and is expected as a reliable tool for the discovery and design of novel materials~\cite{Giustino2007,Giustino2017,Ponce2018,Ponce2020,Ponce2021}.
It has been proven successful in various bulk and \textit{ideal} 2-dimensional (2D) systems, such as GaAs~\cite{Zhou2016,Ma2018,Brunin2020,Ponce2021} and monolayer MoS$_2$~\cite{Guo2019,Song2022}. 
For bulk materials, the systems are typically considered to be infinitely large along all three dimensions, with no boundary effects.
For \textit{ideal} 2D materials, the carrier transport is strictly confined to the 2D plane, assuming no out-of-plane motion or surface scattering.
However, in practical low-dimensional systems, including thin films or nanowires with finite thicknesses or diameters, the carrier motion remains three-dimensional while being subject to surface scattering at the boundaries.
Consequently, the carrier mobility can exhibit a clear spatial gradient along the direction of confinement, resulting in an effective mobility that may significantly deviate from the bulk value.
In previous computational studies~\cite{Kumar2022,VanTroeye2023}, the phenomenological Fuchs-Sondheimer model~\cite{Fuchs1938,Sondheimer1952} was usually applied to estimate the effect of surface scattering.
Recently, a fully \latin{ab initio} method based on the space-dependent Boltzmann transport equation (SD-BTE) has been used to study the electrical conductivity in 2D metal films~\cite{Zhang2024}.
In this work, we generalize the \latin{ab initio} method to study carrier mobility in one-dimensional (1D) semiconductor nanowires.
Both e-ph scattering and surface scattering are considered, and the dependence of mobility on the diameter, orientation, cross-sectional shape, and temperature is explored thoroughly.
Silicon and wurtzite gallium nitride are chosen to be two representative examples for the application of our computational framework, as they cover different polarities and crystal symmetries.

The carrier mobility is calculated by~\cite{Ponce2018,Ponce2020}
\begin{align}\label{eq:mob}
    \mu_{\alpha\beta} = \frac{-1}{V_\mathrm{uc}n_\mathrm{c}}\sum_n \int_\mathrm{BZ} \frac{\mathrm{d}\mathbf{k}}{\Omega_\mathrm{BZ}} v_{n\mathbf{k}\alpha} \partial_{E_\beta} f_{n\mathbf{k}},
\end{align}
where $\alpha$ and $\beta$ are Cartesian directions, $V_\mathrm{uc}$ the volume of the unitcell, $n_\mathrm{c}$ the carrier concentration, $n$ the band index, $\mathbf{k}$ the electronic wavevector, $\Omega_\mathrm{BZ}$ the first Brillouin zone volume, $v$ the band velocity, $E$ the electric field, $f$ the Fermi-Dirac occupation function, and $\partial_{E_\beta} f_{n\mathbf{k}} \equiv (\partial f_{n\mathbf{k}} / \partial E_\beta) \vert _{\mathbf{E=0}}$.
For ideal bulk materials, the linearized Boltzmann transport equation (BTE) reads~\cite{Ponce2018,Ponce2020}
\begin{align}\label{eq:bte}
    \partial_{E_\beta}f_{n\mathbf{k}} & = ev_{n\mathbf{k}\beta}\frac{\partial f_{n\mathbf{k}}^0}{\partial \varepsilon_{n\mathbf{k}}}\tau_{n\mathbf{k}} \\ \notag &+ \tau_{n\mathbf{k}} \sum_m \int_\mathrm{BZ} \frac{\mathrm{d}\mathbf{q}}{\Omega_\mathrm{BZ}}\Gamma_{m\mathbf{k+q}\rightarrow n\mathbf{k}} \partial_{E_\beta}f_{m\mathbf{k+q}},
\end{align}
where $e$ is the elemental charge, $f^0$ the Fermi-Dirac occupation without external field, $\varepsilon$ the electronic eigenvalue, $\tau$ the scattering lifetime (inverse of the total scattering rate), $m$ the band index of the final state, $\mathbf{q}$ the phonon wavevector, and $\Gamma$ the transition rate.

In finite-size systems, the transport property depends on the spatial position $\mathbf{r}$, and the SD-BTE becomes
\begin{align}\label{eq:sdbte}
    [ & 1 + \tau_{n\mathbf{k}}\mathbf{v}_{n\mathbf{k}}\cdot\nabla ] \partial_{E_\beta}f_{n\mathbf{k}}(\mathbf{r}) = ev_{n\mathbf{k}\beta}\frac{\partial f_{n\mathbf{k}}^0}{\partial \varepsilon_{n\mathbf{k}}}\tau_{n\mathbf{k}} \\ \notag &+ \tau_{n\mathbf{k}} \sum_m \int_\mathrm{BZ} \frac{\mathrm{d}\mathbf{q}}{\Omega_\mathrm{BZ}}\Gamma_{m\mathbf{k+q}\rightarrow n\mathbf{k}} \partial_{E_\beta}f_{m\mathbf{k+q}}(\mathbf{r}).
\end{align}
Both eqs~\ref{eq:bte} and \ref{eq:sdbte} can be solved exactly through iteration, or approximately based on the self-energy relaxation time approximation (SERTA), which neglects the second term on the right-hand side.
The carrier mobility is then obtained as a function of spatial position, \latin{i.e.}, $\mu(x,y)$ for 1D nanowires, where $(x, y)$ is the coordinate on the cross-section.
The effective value of mobility can be calculated by averaging over the cross-section, yielding as a scalar
\begin{align}
    \mu = \frac{\int_S \mu(x,y)\mathrm{d}x\mathrm{d}y}{\int_S \mathrm{d}x\mathrm{d}y},
\end{align}
where $S$ denotes the domain of the cross-section.
The detailed formulations of the theoretical framework are provided in Sec.~S1 of the Supporting Information (SI).

In this work, we employ diffusive scattering as the boundary condition to assess the strongest effect of surface scattering on carrier mobility.
It assumes that the carrier distributions restore their equilibrium states after surface scattering~\cite{Zhang2024}, thereby providing a theoretical reference for carrier mobility in the surface-roughness-dominated regime.
Given that real surfaces exhibit considerable complexities (\latin{e.g.} atomic-scale roughness, defects, and chemisorption), this boundary condition is a reasonable assumption with a clear physical interpretation and computational tractability within an \latin{ab initio} framework, setting a fundamental benchmark for understanding the size effect on the charge transport property in 1D nanostructures.

\begin{figure*}[tb]
    \centering
    \includegraphics[width=\linewidth]{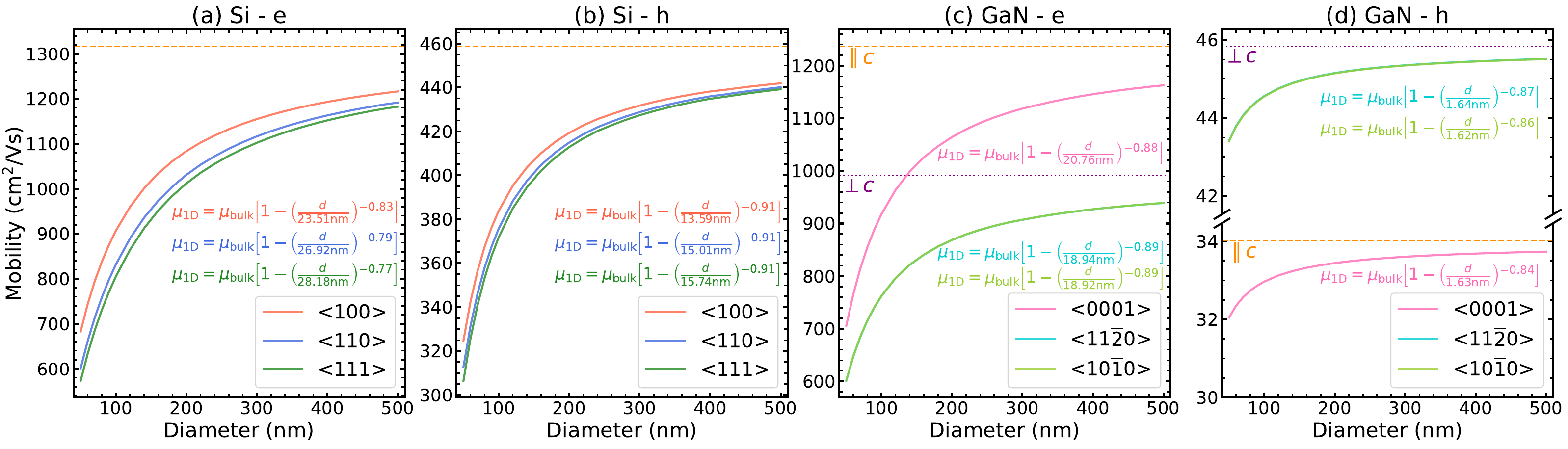}
    \caption{Diameter-dependent room-temperature carrier mobility in nanowires with circular cross-sections and different crystallographic orientations. The horizontal dashed or dotted lines denote the corresponding bulk mobility values.}
    \label{fig:orientation}
\end{figure*}

\begin{figure*}
    \centering
    \includegraphics[width=\linewidth]{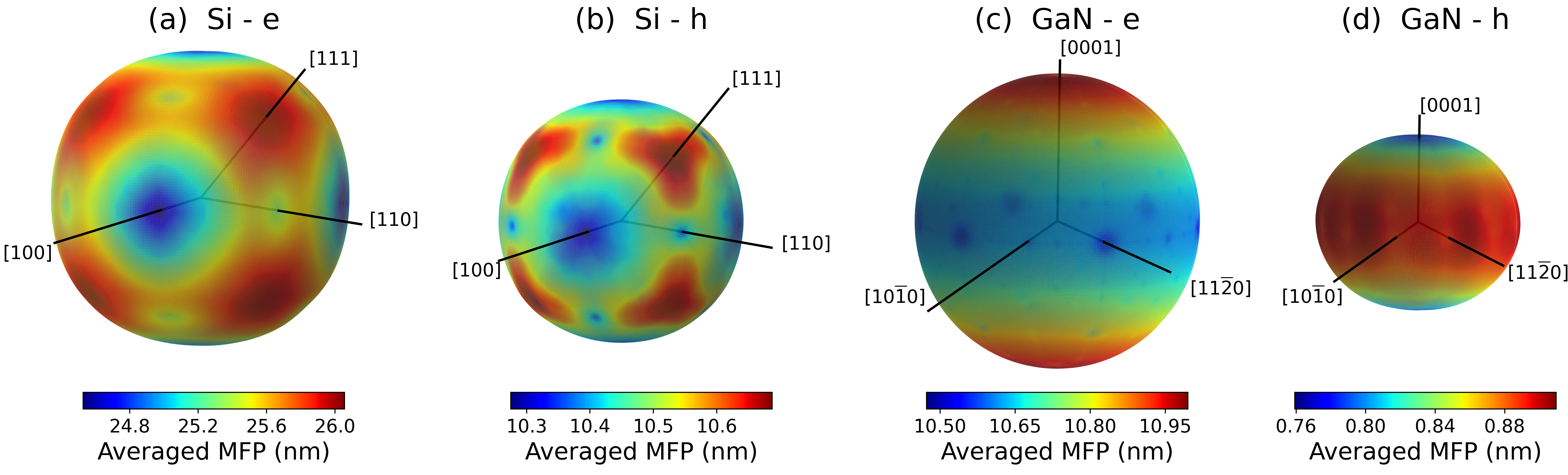}
    \caption{Three-dimensional visualization of direction-dependent carrier MFP at room temperature. Each point on the surface corresponds to a crystallographic direction, with the radial distance and color indicating the averaged MFP weighted by carrier density.}
    \label{fig:mfp}
\end{figure*}

We first calculated the electron-phonon scattering in bulk Si and GaN.
Electronic eigenvalues were calculated using the $G_0W_0$ method to obtain the quasiparticle band structure~\cite{yambo2009,yambo2019}.
Phonon eigenmodes and e-ph matrix elements were calculated based on density-functional perturbation theory (DFPT)~\cite{Gonze1997a,Gonze1997b}.
These quantities were interpolated unto ultrafine momentum grids based on maximally localized Wannier functions (MLWFs)~\cite{Giustino2007}, and the carrier scattering rates were calculated self-consistently~\cite{Lihm2024}.
Then the phonon-limited bulk mobilities were obtained by solving eq~\ref{eq:bte}.
The computational details are provided in Sec.~S2.
For both Si and GaN, the calculated mobilities based on BTE are in good agreement with previous experimental values, as compared in Table~S2.
Regarding SERTA, although it can yield reasonable mobility results for Si, it severely underestimates the mobilities of GaN by more than 30\%.
This can be attributed to the significant piezoelectric scattering and Fr\"ohlich interaction in GaN, both of which are mediated by long-wave phonons.
Therefore, small-angle forward scattering dominates in GaN, leading to the failure of SERTA.
This tendency has also been observed in other polar systems~\cite{Ma2018,Claes2022,He2025}.
The detailed analysis of the scattering mechanisms in Si and GaN is provided in Sec.~S3.

\begin{figure*}[tb]
    \centering
    \includegraphics[width=\linewidth]{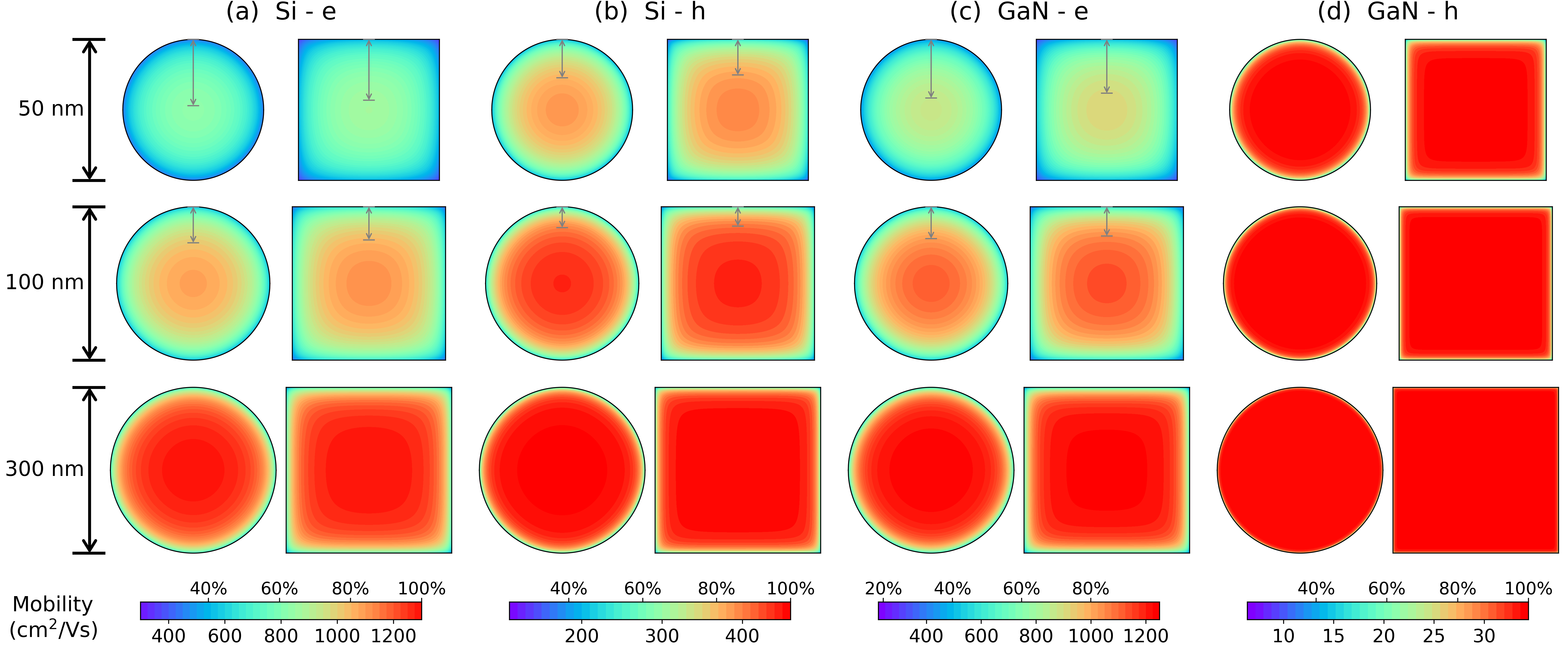}
    \caption{Cross-sectional distribution of room-temperature carrier mobility for nanowires with different diameters and cross-sectional shapes. The silver arrows indicate the fitted $d_0$ based on eq~\ref{eq:mob-diameter} (not shown if $d_0 \ll d$). Crystallographic orientations are $<$100$>$ for Si and $<$0001$>$ for GaN. Percentages are with respect to the bulk value.}
    \label{fig:cross-section}
\end{figure*}

Figure~\ref{fig:orientation} presents the carrier mobility of Si and GaN nanowires as a function of diameter. 
As expected, the effective mobility increases with diameter, which can be attributed to the reduced surface-to-volume ratio and thus a weaker contribution of surface scattering.
On the other hand, the deviation of mobility in nanowires from the bulk value varies considerably across materials.
This difference stems from the relative importance of surface scattering, which depends on the diameter compared with the carrier mean free path (MFP, as shown in Figure~\ref{fig:mfp}).
For example, for nanowires with a diameter of 100~nm, the hole mobility of GaN recovers 97\% of its bulk value, whereas the electron mobility of Si is below 70\% of the corresponding bulk value, as the MFP of electrons in Si is an order of magnitude greater than that of holes in GaN (Figure~\ref{fig:mfp}).

Furthermore, the data in Figure~\ref{fig:orientation} indicate that carrier mobility $\mu_\mathrm{1D}$ in nanowires follows a systematic dependence on diameter $d$, which can be quantitatively described by the following expression
\begin{equation}\label{eq:mob-diameter}
    \mu_\mathrm{1D} = \mu_\mathrm{bulk}\left[1-\left(\frac{d}{d_0}\right)^{-\beta}\right]\quad(d>d_0),
\end{equation}
where $\mu_\mathrm{bulk}$ denotes the bulk mobility, and $d_0$ and $\beta$ are system-specific parameters.
Eq~\ref{eq:mob-diameter} can be transformed into an equivalent expression that emphasizes the physical interpretation
\begin{equation}\label{eq:mob-surf}
    \mu_\mathrm{s} = \mu_\mathrm{bulk}\left[\left(\frac{d}{d_0}\right)^\beta-1\right],
\end{equation}
where $\mu_\mathrm{s}$ defines the mobility that solely includes the effect of surface scattering, which follows Matthiessen's rule
\begin{equation}\label{eq:matthiessen}
    \frac{1}{\mu_\mathrm{s}} = \frac{1}{\mu_\mathrm{1D}} - \frac{1}{\mu_\mathrm{bulk}}.
\end{equation}
A linear fit of the transformed variables (Figure~S4) is performed to obtain both parameters, which are summarized in Figure~\ref{fig:orientation}.
It is found that $d_0$ is comparable in magnitude to the averaged MFP shown in Figure~\ref{fig:mfp}.
To clarify the physical significance of $d_0$, we visualize the local distribution of carrier mobility across the nanowire cross-section in Figure~\ref{fig:cross-section}.
The results indicate that $d_0$ defines the thickness of a boundary layer, within which the local mobility is significantly lower than the bulk value and exhibits an appreciable spatial gradient, which is consistent with the boundary condition of diffusive scattering.
As the diameter increases, the relative contribution of the boundary layer weakens, leading to a larger $d/d_0$ ratio and higher mobility.

\begin{figure*}[tb]
    \centering
    \includegraphics[width=\linewidth]{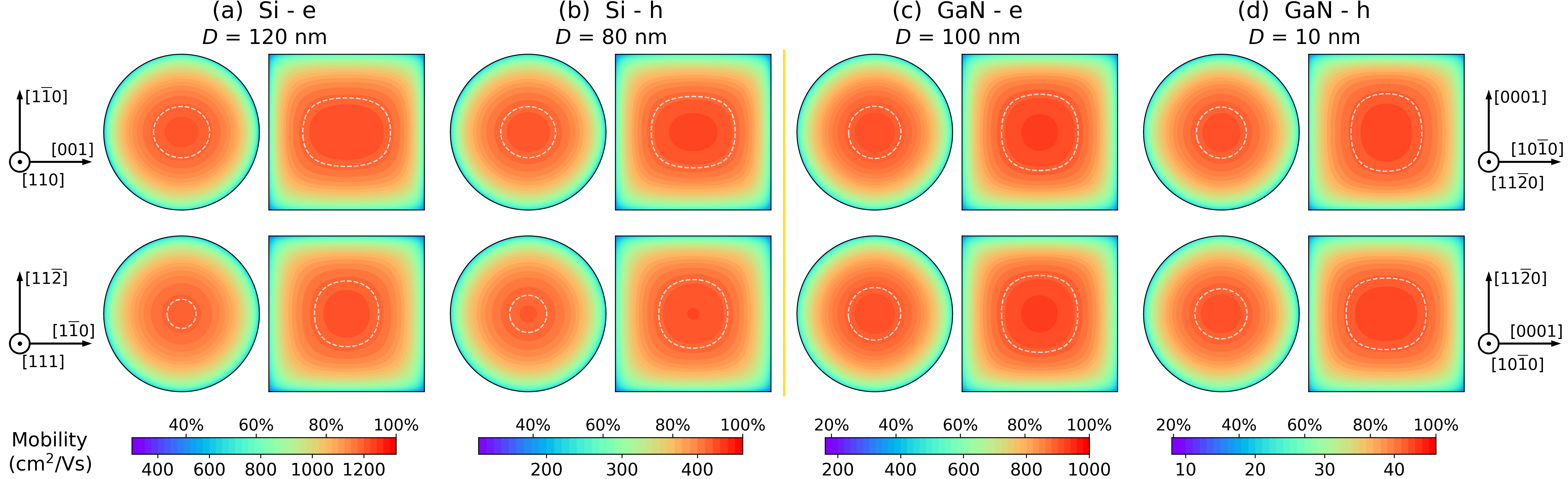}
    \caption{Cross-sectional distribution of room-temperature carrier mobility for nanowires with different crystallographic orientations and cross-sectional shapes. The white dashed lines represent the contour of $\mu = 0.9\mu_\mathrm{bulk}$. Percentages are with respect to the bulk value.}
    \label{fig:cross-section-orientation}
\end{figure*}

We further compare the mobility of nanowires with circular and square cross-sections.
As shown in Figure~\ref{fig:cross-section}, square cross-sections with a side length equal to the diameter of circular ones have a lower minimum local mobility, which is located at the corner, possibly due to the strong spatial confinement near the right angles.
On the other hand, square cross-sections show a higher maximum of local mobility.
Overall, square cross-sections yield a higher effective mobility (Figure S5), possibly due to their larger cross-sectional area.
Linear fitting based on eq~\ref{eq:mob-diameter} reveals little difference in $\beta$ between the two shapes, whereas $d_0$ is larger for square cross-sections (Table~S3).
Additionally, for the materials investigated here, it is also found that $\frac{\pi}{4}d_{0,\mathrm{square}}^2 > d_{0,\mathrm{circular}}^2$.
Therefore, nanowires with a square cross-section generally have a lower mobility than their circular counterparts with the same cross-sectional area, consistent with the calculated results in Figure~S6.

Mobility also depends on nanowire orientation (Figure~\ref{fig:orientation}).
The fitted parameters listed in Table~S3 indicate that both $\beta$ and $d_0$ may vary with orientation.
For Si, the bulk carrier mobility is isotropic because of its high crystal symmetry.
Nonetheless, its $<$100$>$, $<$110$>$, and $<$111$>$ oriented nanowires have distinct mobilities, as surface scattering depends on the MFP within the cross-sectional plane.
The anisotropy of the band velocity leads to different in-plane MFPs for these orientations, resulting in different surface scattering rates and thus distinct mobilities.
For GaN, its anisotropy leads to a more pronounced directional difference.
For electrons, the phonon-limited mobility parallel to the $c$-axis is higher than that perpendicular to the $c$-axis (Table~S2), and the smaller in-plane MFPs also lead to reduced surface scattering in $<$0001$>$ oriented nanowires.
A similar tend holds for holes in GaN, where the mobility along the $<$0001$>$ direction is much lower.
In contrast, mobilities in $<$11$\overline{2}$0$>$ and $<$10$\overline{1}$0$>$ oriented nanowires are almost identical , consistent with the similar in-plane MFP values.

Figure~\ref{fig:cross-section-orientation} visualizes the mobility distribution across cross-sections of differently oriented nanowires.
A key difference from Figures~\ref{fig:cross-section} is that the cross-sections rendered in Figure~\ref{fig:cross-section} lie in high-symmetry \{100\} and \{0001\} crystallographic planes for Si and GaN, respectively.
Consequently, the contours of local mobility are a series of concentric circles for circular cross-sections, and have fourfold rotational symmetry for square cross-sections.
In Figure~\ref{fig:cross-section-orientation}, this symmetry is broken.
The spatial gradient of local mobility is more significant along the direction with shorter MFPs, \latin{e.g.}, $<$0001$>$ direction for holes in GaN compared with $<$10$\overline{1}$0$>$ and $<$11$\overline{2}$0$>$ directions.
This suggests that in strongly anisotropic systems, tailoring the aspect ratio of the cross-section according to the MFP anisotropy could enhance carrier mobility without changing the cross-sectional area.

\begin{figure*}[tb]
    \centering
    \includegraphics[width=\linewidth]{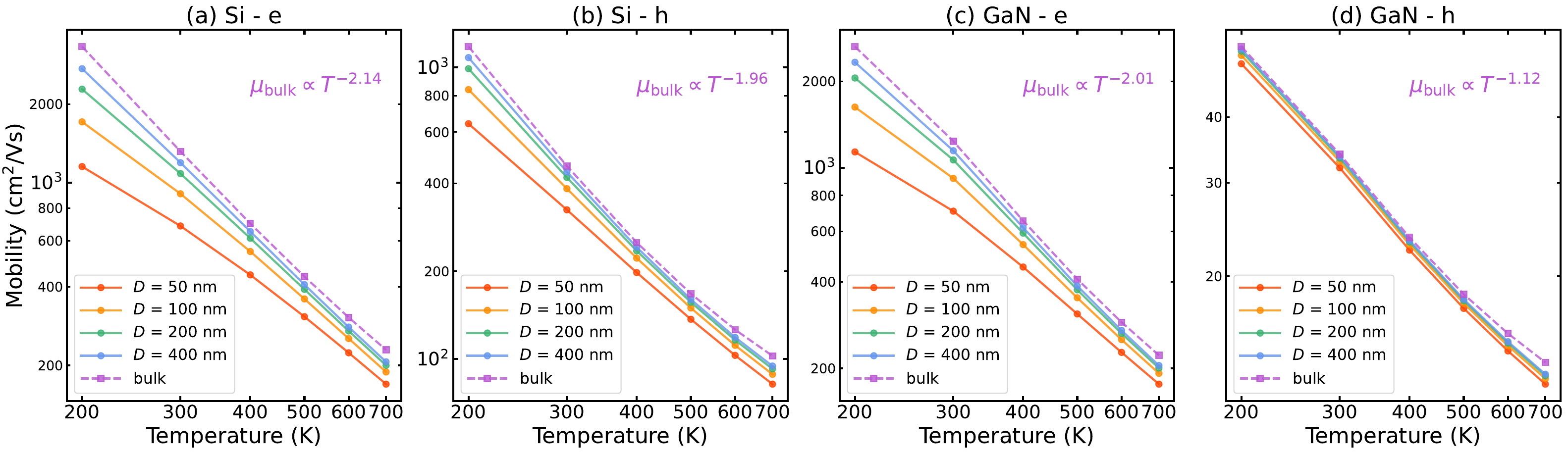}
    \caption{Temperature-dependent carrier mobilities in circular nanowires with selected diameters. Crystallographic orientations are $<$100$>$ for Si and $<$0001$>$ for GaN. The results of bulk materials are provided for comparison, which are fitted to the power law $\mu \propto T^{-\alpha}$.}
    \label{fig:temp}
\end{figure*}

\begin{table*}
    \centering
    \begin{tabular}{cccccc}
      \hline
         System & Orientation & Diameter (nm) & $\mu$, this work (\mobunit) & $\mu$, exp. (\mobunit) & Ref \\
         \hline
         \multirow{2}{*}{Si-h} & $<$110$>$ & 20 & 208                       & 130                    & \citenum{Gunawan2008} \\
                & $<$111$>$   & 20            &  201                        & 119                    & \citenum{Duan2003} \\
         GaN-e  & $<$0001$>$  & 310           &  1122                       & 820$\pm$120            & \citenum{Parkinson2012} \\
         GaN-h  & $<$0001$>$  & 20            &  29                         & 12                     & \citenum{Zhong2003} \\
      \hline
    \end{tabular}
    \caption{Comparison of calculated and experimental room-temperature carrier mobility in nanowires.}
    \label{tab:mob}
\end{table*}

Carrier mobility is also a function of temperature.
In bulk materials, phonon-limited mobility typically decreases as $\mu\propto T^{-\alpha}$ over a temperature range where the predominant scattering mechanism remains unchanged~\cite{Ponce2021,He2025}, due to increased phonon occupation.
This power-law relationship is also observed in the bulk materials investigated here (purple curves in Figure~\ref{fig:temp}).
To understand the temperature dependence of carrier mobility in nanowires, we fitted the parameters $\beta$ and $d_0$ in eq~\ref{eq:mob-diameter} at a series of temperatures, as listed in Table~S3. 
It is found that $d_0$ decreases as temperature increases, consistent with the smaller MFP resulting from enhanced e-ph scattering.
Meanwhile, $\beta$ also decreases with increasing temperature.
Therefore, the function $\mu_\mathrm{1D} (T) $ is likely to take a more complex form.
As seen in Figure~\ref{fig:temp}, the temperature dependence of mobility in nanowires deviates notably from that of bulk materials.
At low temperatures, where the diameter is small relative to the MFP, mobility becomes less sensitive to temperature variation.
As the temperature increases, the MFP shortens significantly due to enhanced e-ph scattering, and the temperature dependence of mobility gradually recovers the bulk trend.
Overall, across a broad temperature range, the temperature dependence of mobility in nanowires cannot be accurately described by a single power-law relationship with a single exponent $\alpha$.
For example, the red curve in Figure~\ref{fig:temp}a exhibits non-negligible deviation from linearity in the log-log plot, which can be attributed to a transition in the predominant scattering mechanism from surface scattering (\latin{e.g.}, $T$ = 200~K, where surface-limited mobility $\mu_\mathrm{s}$ = 1767~\mobunit $<$ phonon-limited mobility $\mu_\mathrm{bulk}$ = 3317~\mobunit) to e-ph scattering (\latin{e.g.}, $T$ = 400~K, where $\mu_\mathrm{s}$ = 1220~\mobunit $>$ $\mu_\mathrm{bulk}$ = 698~\mobunit).

To connect our theoretical model to practical applications, we compare our calculated mobility with experimental data from the literature for various nanowires, as listed in Table~\ref{tab:mob}.
It is found that our predicted results are systematically higher than those measured in experiments.
We attribute this discrepancy to multiple critical differences between our theoretical framework and real-world conditions.
Firstly, our model assumes uniform carrier distribution across the cross-section.
In contrast, experimental data are typically extracted from measurements based on field effect transistor devices, where the gate electric field confines carriers near the surface.
This redistribution of carriers enhances the impact of surface scattering, resulting in lower mobility.
Secondly, experimentally synthesized nanowires usually contain structural imperfections such as dislocations, charged impurities, and surface states, which can serve as additional scattering sources.
Thirdly, there may be other extrinsic factors introduced by the measurement method, such as contact resistance, which can also lead to considerable underestimation of carrier mobility.

Our theoretical model, which quantifies e-ph and surface scattering within an ideal lattice, is inherently free from uncertainties caused by sample quality, measurement limitations, and environment fluctuations.
It is thus expected to provide a reliable benchmark for intrinsic mobility in nanowire systems.
The gap between theoretical predictions and experimental data can serve as a quantitative measure of the effect of crystal imperfection introduced during synthesis and the methodological limitations of electrical transport measurement.
Therefore, this theoretical model provides not only a predictive tool, but also a useful reference for evaluating the performance of nanowire systems and guiding the optimization of both material synthesis and device engineering.

Finally, we briefly discuss the differences between solving the BTE for bulk materials and SD-BTE for nanowires.
Unlike the spatially homogeneous BTE, SD-BTE for nanowires typically involves a series of partial differential equations that generally require real-space discretization for numerical solution.
Hence, the choice of grid distribution can significantly influence the convergence with respect to grid size.
We performed convergence tests for square cross-sections using both uniform and non-uniform orthogonal grids generated using a cosine function, with the latter providing enhanced resolution near the boundary.
As shown in Figure~S7, non-uniform grids achieve faster convergence, particularly when the side length is much larger than the MFP, since the gradient distribution is obvious only in a narrow region near the surface (see \latin{e.g.} the third row of Figure~\ref{fig:cross-section}c).
For circular cross-sections, non-uniformed orthogonal grids based on a cosine function can ensure sufficient grid points on the cross-sectional boundary compared with uniform orthogonal grids, and also avoid the singularity issue associated with polar grids.
Therefore, non-uniform orthogonal grids are recommended for both rectangular and circular cross-sections.
Figure~S7 also shows that the grid size needed for convergence decreases as side length increases, possibly due to the diminishing contribution of the boundary layer.

Another important consideration is the applicability of SD-SERTA.
As shown in Figure~S8, the relative underestimation of mobility by SD-SERTA for nanowires is not the same as that by BTE for the bulk material, but a function of diameter.
For Si, where the SERTA and BTE results for bulk are in close agreement, SD-SERTA also provides reasonable estimations.
Nevertheless, for GaN, where SERTA severely underestimates the bulk mobility, the accuracy of SD-SERTA depends strongly on diameter.
As the diameter decreases toward the MFP, SD-SERTA results gradually approach the SD-BTE values.
The possible reason is that the boundary condition of diffusive scattering is largely consistent with the assumption of SERTA and SD-SERTA that carriers completely lose their momentum after each scattering event.
In the large-diameter limit, the carrier mobility recovers the bulk value, and the underestimation by SD-SERTA matches its bulk counterpart.
For intermediate diameters, however, the accuracy of SD-SERTA may be less predictable.
Therefore, although SD-SERTA may be applicable in some specific cases, iterative solutions to SD-BTE are generally required for accurate results.

In summary, we have developed a theoretical framework that integrates bulk e-ph scattering and surface scattering to predict carrier mobility in 1D semiconductor nanowires.
Using an iterative solver for the SD-BTE, we apply this method to Si and GaN with diffusive scattering as the boundary condition.
The calculated mobility is found to be a function of diameter, which reads $\mu_\mathrm{1D} = \mu_\mathrm{bulk} \left[ 1 - \left(d/d_0\right)^{-\beta} \right]$, with $d_0$ and $\beta$ as system-specific parameters.
$d_0$ defines the thickness of a boundary layer where the local mobility exhibits an appreciable gradient distribution, and is comparable to the carrier MFP.
Both parameters jointly describe the competition between e-ph scattering and surface scattering.
The cross-sectional shape mainly affects $d_0$, while the crystallographic orientation has a considerable effect on both $d_0$ and $\beta$ in systems with significant MFP anisotropy.
At elevated temperature, both $d_0$ and $\beta$ decrease, due to the reduced MFP and altered balance in scattering mechanisms. 
The temperature dependence of mobility in nanowires may deviate from the bulk power-law relationship, which is attributed to the transition from the surface-dominated regime to the electron-phonon-dominated regime as temperature increases.
Compared with our predicted values, experimental data are generally lower.
This gap quantifies the impact of extrinsic factors under real-world experimental conditions such as structural imperfections and measurement limitation.
This \latin{ab initio} method thus provides a predictive tool for unexplored materials and also an intrinsic performance benchmark for known systems, offering a reference for experimental optimization.

\begin{acknowledgement}

This work was supported by research funds from Shanghai Advanced Silicon Technology Co., Ltd. and also the Natural Science Foundation of Shanghai (Grant No. 23ZR1403300).

\end{acknowledgement}

\begin{suppinfo}

Theoretical framework, computational details, transport properties in bulk materials, additional figures and table.

\end{suppinfo}

\bibliography{ref}

\end{document}